\documentclass[reprint,superscriptaddress, amsmath,amssymb,aps,prb]{revtex4-2}

\usepackage{graphicx,dcolumn,bm,hyperref,gensymb,xcolor}
\begin{document}

\title{{Indirect} Magnetoelectric Coupling via Skew Scattering by Orbital Angular Momentum} 



\author{Adam B. Cahaya}
\email[]{adam@sci.ui.ac.id}
\affiliation{Department of Physics, Faculty of Mathematics and Natural Sciences, Universitas Indonesia, Depok 16424, Indonesia}

\date{\today}

\begin{abstract}
Recent experimental observations of exchange bias in the La$_{0.67}$Sr$_{0.33}$MnO$_{3}$/LaAlO$_{3}$/SrTiO$_{3}$ heterostructure, which lacks an intrinsic antiferromagnetic layer, have sparked theoretical investigations into the underlying mechanisms. While traditional theories suggest that exchange bias in spin valve structures is mediated by conduction electrons in metallic spacers, the transport properties of LaAlO$3$ are dominated by its valence electrons, raising new questions about the origin of this phenomenon. In this work, we propose a theoretical model where the electronic band structure of LaAlO$_3$ is treated as a valence band perturbed by skew scattering, which is sensitive to orbital angular momentum. Our analysis reveals a significant magnetoelectric effect at the La$_{0.67}$Sr$_{0.33}$MnO$_3$/LaAlO$_3$ interface, which induces a coupling between the interface magnetization and the electric field from two dimensional electron gas at LaAlO$_{3}$/SrTiO$_{3}$ interface. This magnetoelectric coupling is found to drive the observed exchange bias, highlighting the role of electric polarization in influencing the magnetic properties of the heterostructure. 

\end{abstract}


\maketitle 

\section{Introduction}

Efficient manipulation of magnetic moments is crucial for enhancing the data writing efficiency in magnetic memory devices, a key challenge in spintronics \cite{Dieny2020}. The ability to control magnetization with high precision is essential for the development of advanced magnetic storage technologies, including magnetic random-access memory (MRAM) and other spintronic devices. This requires the ability to fix the magnetization of one layer in a bilayer system, which is typically achieved through the phenomenon of exchange bias \cite{C7NR05491B}. 

{Exchange bias manifests when a ferromagnetic layer is exchange-coupled to an adjacent antiferromagnetic layer, resulting in a unidirectional anisotropy that shifts the magnetic hysteresis loop away from zero field \cite{KIWI2001584}. This effect originates microscopically from the exchange interactions between ferromagnetic spins and uncompensated antiferromagnetic spins at the interface, which pin the ferromagnetic magnetization \cite{Stamps_2000}. The complex spin structures at the interface, including spin canting, domain formation, and magnetic frustration, strongly influence the magnitude and stability of the exchange bias \cite{NOGUES1999203, Berkowitz1999, Nogues2005, KIWI2001584}. Previous studies show interfacial magnetic disorder and roughness, as described in random-field models, generate irreversible spin configurations that are critical for the exchange bias \cite{Malozemoff1987}. 

Early theoretical work demonstrated that the precise interfacial spin alignment can lead to spin-flop coupling phenomena that critically determine direction and magnitude of the exchange bias \cite{Schulthess1998, Koon1997}. Such studies demonstrate how noncollinear spin configurations and domain wall formation in the antiferromagnet near the interface stabilize the exchange bias effect. Experimentally, oxide heterostructures have served as excellent platforms to probe and engineer these interfacial magnetic interactions. Investigations on ferrite-based multiferroic heterostructures revealed electric-field controllable exchange bias, where manipulation of antiferromagnetic domain states by electric polarization enables reversible tuning of the bias \cite{Bea2006, Bea2008}. Moreover, studies on manganite-based heterostructures have highlighted the role of epitaxial strain and interfacial coupling in modulating magnetic order and exchange bias, confirming the sensitivity of the effect to structural and electronic interface characteristics \cite{Dho2006}.

Beyond classical exchange bias, recent advances show that electric fields can modulate exchange bias via magnetoelectric coupling, offering energy-efficient control of magnetization \cite{Chu2008, Martin2008, Dong2009}. In multiferroic and complex oxide heterostructures, electric-field-driven rearrangement of antiferromagnetic domains induces changes in exchange bias, thus linking electric polarization with magnetic order \cite{Padhan2005, Choi2007}. This coupling paves the way for spintronic devices that combine ferroelectric and magnetic functionalities with low power consumption \cite{PhysRevLett.110.067202, CHEN2022169753, SONG201733, Nozaki_2012}.

Interestingly, unconventional exchange bias effects have been observed in systems lacking a classical antiferromagnetic layer. For instance, in La$_{0.67}$Sr$_{0.33}$MnO$_3$/LaAlO$_3$/SrTiO$_3$ (LSMO/LAO/STO) heterostructures, exchange bias phenomena have been attributed to interfacial spin-orbit coupling and broken inversion symmetry rather than traditional antiferromagnetic pinning \cite{Lu2016, LEE2002264, 10145485}. The magnetoelectric coupling in these systems is thought to be mediated by conduction electrons within the oxide layers, with spin-orbit interactions providing the key link between magnetic and electric degrees of freedom \cite{Padhan2005, Choi2007}. These findings extend the conventional framework of exchange bias and suggest alternative pathways to control magnetic states in oxide interfaces.

However, prior theoretical models often simplify the electronic structure of oxide layers as metallic with free-electron-like bands, which does not reflect the true band structure of materials like LAO, where the Fermi level lies within the valence band and an indirect gap exists between valence and conduction bands \cite{anderson2023}. Such discrepancies indicate that more realistic models must account for the actual electronic dispersion and scattering mechanisms. In particular, skew scattering arising from small but finite orbital angular momentum at the LSMO/LAO interface has been proposed as a mechanism enabling coupling between orbital moments and charge density \cite{Cahaya2021}, thereby enhancing magnetoelectric interactions that can induce exchange bias .
}

The rest of this article is organized as follows. In Sec.\ref{SecLinearResponse}, we explore the charge density induced by skew scattering at the LSMO/LAO interface, focusing on the role of orbital magnetic moments. In Sec.\ref{SecResult}, we demonstrate how this electric polarization, enabled by the magnetoelectric effect, induces exchange bias. Finally, in Sec.~\ref{SecSummary}, we summarize our findings.
By refining the microscopic understanding of exchange bias in magnetic heterostructures, we hope to contribute to the development of more efficient and scalable spintronic devices that could revolutionize data storage and processing technologies.

{

\section{Theoretical model}
\label{SecLinearResponse}

To study the indirect magnetoelectric coupling in the LSMO$|$LAO$|$STO heterostructure, we consider the following model: a system of itinerant charges in the spacer (LAO) perturbed by the magnetic field originating from local spins at the LSMO interface and the local electric field at the STO interface. These local fields polarize the itinerant charges, which in turn mediate an indirect magnetoelectric coupling.

The electronic band structure of LAO, illustrated in Fig.~\ref{Fig.LAO}, is determined using density functional theory (DFT) \cite{BANDscm,PhysRevB.44.7888,Kadantsev2007}. From the band structure, we find that the Fermi energy of LAO lies at its valence band, with a peak located at the R-point. The conduction band begins at the $\Gamma$-point, indicating an indirect band gap. This is consistent with data from material databases \cite{MatProj,Benam2014,Jain2013}. As a result, the indirect magnetoelectric coupling is mediated by the valence band. 

We model the unperturbed system with a parabolic dispersion, given by the Hamiltonian:
\begin{align}
H_0 = \sum_{\mathbf{k}\alpha} \varepsilon_{\mathbf{k}} a_{\mathbf{k}\alpha}^\dagger a_{\mathbf{k}\alpha},\ 
\varepsilon_{\mathbf{k}} = -\frac{\hbar^2 }{2m}(\mathbf{k} - \mathbf{p})^2.
\end{align}
The Fermi energy is \(\varepsilon_F \equiv -\hbar^2 k_F^2/2m\). The band maximum occurs at momentum $\hbar \textbf{p}$, corresponding to the R-point. The effective mass is estimated to be $m\approx 92m_e$. 

\begin{figure}[t]
\includegraphics[width=\columnwidth]{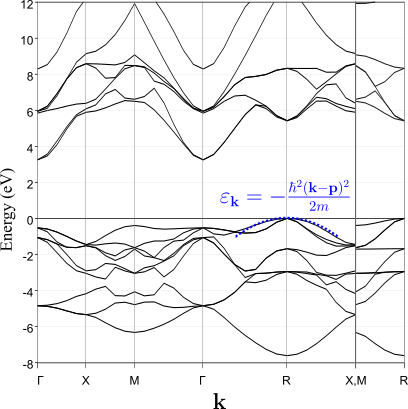}
\caption{ Band structure of LAO from DFT calculation \cite{BANDscm,PhysRevB.44.7888,Kadantsev2007}, indicating a $p$-type character, in agreement with existing material database \cite{MatProj,Benam2014,Jain2013}.}
\label{Fig.LAO}
\end{figure}

}
At the interface between LSMO and LAO, the localized moments of Mn influence the electrons in LAO through two primary mechanisms: exchange interaction \(H_{\rm J\cdot s}\) and skew scattering \(H_{\rm J\cdot l}\) \cite{Kondo1962,Fert1976}. The total interaction Hamiltonian is given by 

\begin{align}
H_{\rm int} =& H_{\rm J\cdot s} + H_{\rm J\cdot l}, \label{Eq.Hamiltonian}
\end{align}
where the exchange interaction \(H_{\rm J\cdot s}\) and skew scattering \(H_{\rm J\cdot l}\) are defined as {\cite{Kondo1962,Fert1976,Cahaya2021}}

\begin{align*}
H_{\rm J\cdot s} = -\mathcal{I} \sum_{\mathbf{kq}\alpha\beta}(g-1)\mathbf{J} \cdot a_{\mathbf{k}\alpha}^\dagger \boldsymbol{\sigma}_{\alpha\beta} a_{\mathbf{q}\beta},\\
H_{\rm J\cdot l} = -i\mathcal{F} \sum_{\mathbf{kq}\alpha}(2-g)\mathbf{J} \cdot (\mathbf{q} \times \mathbf{k}) a_{\mathbf{k}\alpha}^\dagger a_{\mathbf{q}\alpha}, 
\end{align*}
where \(\boldsymbol{\sigma} = [\sigma_x \ \sigma_y \ \sigma_z]\) represents the Pauli matrices, \(\mathbf{J}\) represents the localized angular moment and \(g\) is the g-factor of Mn, which can deviate from 2 \cite{Cherif2015}. The constants \(\mathcal{I}\) and \(\mathcal{F}\) represent the exchange and skew-scattering strengths, respectively. {
$H_{\rm J\cdot l}$ is order of magnitude smaller than $H_{\rm J\cdot l}$, with $\mathcal{F}\approx\mathcal{I}/(4k_F^2)$ \cite{Cahaya2021}.} A deviation of \(g\) from 2 implies that the orbital angular momentum \(\mathbf{L} = (2-g)\mathbf{J}\) is not fully quenched. 

\subsection{Linear response theory}

The interactions induce polarization in the charge and spin densities of LAO, which are expressed as
\begin{align}
\rho(\mathbf{r}) =& e \sum_{\mathbf{kq}\alpha} e^{i \mathbf{q} \cdot \mathbf{r}} a_{\mathbf{k}+\mathbf{q}\alpha}^\dagger a_{\mathbf{k}\alpha}, \notag \\
\mathbf{s}(\mathbf{r}) =& \sum_{\mathbf{kq}\alpha\beta} e^{i \mathbf{q} \cdot \mathbf{r}} a_{\mathbf{k}+\mathbf{q}\alpha}^\dagger \boldsymbol{\sigma}_{\alpha\beta} a_{\mathbf{k}\beta}.
\end{align}
The response of the charge and spin densities to the interactions is calculated at linear order using the Kubo formula

\begin{align}
\rho(\mathbf{r}, t) =& \frac{1}{i\hbar} \int_{-\infty}^t dt' \left< \left[ \rho(\mathbf{r}, t), H_{\rm int}(t') \right] \right>, \notag \\
\mathbf{s}(\mathbf{r}, t) =& \frac{1}{i\hbar} \int_{-\infty}^t dt' \left< \left[ \mathbf{s}(\mathbf{r}, t), H_{\rm int}(t') \right] \right>.
\end{align}

The next sections evaluate the spin and charge densities in response to a localized moment \(\mathbf{J}\) placed at the origin.

\subsubsection{Spin Density}
\label{SubSec:spin}

Defining the exchange interaction as
\begin{align}
H_{\rm J\cdot s}
\equiv&-\int d\textbf{r}\ \textbf{m}\cdot\textbf{B}\notag,
\end{align}
\(H_{\rm J \cdot s}\) can be interpreted as an effective magnetic field, 
\begin{equation}
\mathbf{B} = \frac{(g-1) \mathcal{I}}{\mu_B} \mathbf{J} \delta^3(\mathbf{r}),
\end{equation} 
acting on the itinerant spin density, where \(\mu_B\) is the Bohr magneton. Therefore{, the localized form of a dipolar magnetic field \cite{10145485} can be used to estimate $\mathcal{I}= \mu_B^2\mu_0/3\hbar\approx 2.14$ eV. The spin density response is given by \cite{Kittel} }

\begin{align}
\mathbf{s}(\mathbf{r}, t) =& \frac{1}{i\hbar} \int_{-\infty}^t dt' \left< \left[ \mathbf{s}(\mathbf{r}, t), H_{\rm J\cdot s}(t') \right] \right>\notag \\
=& \mu_B\int d\mathbf{r}' \int_{-\infty}^\infty dt' \, \chi_{ij}(\mathbf{r}-\mathbf{r}', t-t') \mathbf{B}(\mathbf{r}', t').
\end{align}
Here, \(\chi_{ij}(\mathbf{r}-\mathbf{r}', t-t')\) is the spin-spin susceptibility, defined as

\begin{align}
\chi_{ij}(\mathbf{r}-\mathbf{r}', t-t') =& \frac{\theta(t-t')}{-i\hbar} \left< \left[ s_i(\mathbf{r}, t), s_j(\mathbf{r}', t') \right] \right>,
\end{align}
and the time evolution of \(\chi_{ij}(\mathbf{q}, t)\) is governed by the Heisenberg equation

\begin{align}
\frac{\partial \chi_{ij}(\mathbf{q}, t)}{\partial t} =& \frac{[\chi_{ij}(\mathbf{q}, t), H_0]}{\partial t} \notag\\
\leftrightarrow \hbar \omega \chi_{ij}(\mathbf{q}, \omega) =& [\chi_{ij}(\mathbf{q}, \omega), H_0].
\end{align}

Within the random phase approximation (RPA), the static spin-spin susceptibility \(\chi_{ij}(\mathbf{r})\) is obtained as

\begin{align}
\chi_{ij}(\mathbf{r}) =& \delta_{ij} \sum_{\mathbf{kq}} e^{i \mathbf{q} \cdot \mathbf{r}} \frac{f_{\mathbf{k}} - f_{\mathbf{k} + \mathbf{q}}}{\varepsilon_{\mathbf{k} + \mathbf{q}} - \varepsilon_{\mathbf{k}}} \equiv \delta_{ij} \chi(\mathbf{r}),
\end{align}
where \(f_{\mathbf{k}}\) is the Fermi distribution function. Expanding this expression in terms of small deviations around the Fermi momentum, we obtain:

\begin{align}
\chi(\mathbf{r}) =& \sum_{\delta\mathbf{k}, \delta\mathbf{q}} e^{i (\delta\mathbf{q} - \delta\mathbf{k}) \cdot \mathbf{r}} \frac{f_{\mathbf{p} + \delta \mathbf{k}} - f_{\mathbf{p} + \delta \mathbf{q}}}{\varepsilon_{\mathbf{p} + \delta \mathbf{q}} - \varepsilon_{\mathbf{p} + \delta \mathbf{k}}}.
\end{align}

After evaluating the integral, we find that \(\chi(\mathbf{r})\) exhibits oscillatory behavior with a spatial frequency of \(2k_F r\):

\begin{align}
\chi(r) =& \frac{m}{8 \hbar^2 \pi^3} \frac{2k_F r \cos(2k_F r) - \sin(2k_F r)}{r^4}. \label{Eq.chiS}
\end{align}
Thus, the spin density around the localized moment \(\mathbf{J}\) is radially symmetric, and can be written as

\begin{align}
\mathbf{s}(\mathbf{r}) =& (g-1)\mathcal{I} \mathbf{J} \chi(r).
\end{align}

\subsubsection{Charge Density}
\label{SubSec:charge}

On the other hand, the Hamiltonian term \(H_{\rm J\cdot l}\) behaves as an effective vector potential \(\mathbf{A}(\mathbf{r})\), which is given by

\begin{equation}
\mathbf{A}(\mathbf{r}) = \frac{(2-g)m\mathcal{F}}{e\hbar} \nabla \times \mathbf{J} \delta^3(\mathbf{r}).
\end{equation}
This vector potential interacts with the current density \(\mathbf{j}(\mathbf{r})\), as described by the following Hamiltonian term

\begin{equation}
H_{\rm J\cdot l} = - \int d\mathbf{r}\, \mathbf{j}(\mathbf{r}) \cdot \mathbf{A}(\mathbf{r}),
\end{equation}
with the current density \(\mathbf{j}(\mathbf{r})\) expressed as

\begin{equation}
\mathbf{j}(\mathbf{r}) = \frac{e\hbar}{m} \sum_{\mathbf{k}, \mathbf{q}, \alpha} e^{i\mathbf{q}\cdot \mathbf{r}} \left( \mathbf{k} + \frac{\mathbf{q}}{2} \right) a_{\mathbf{k} + \mathbf{q}, \alpha}^\dagger a_{\mathbf{k}, \alpha}.
\end{equation}

To describe the response of the charge density \(\rho(\mathbf{r}, t)\) to this interaction, we begin with the general expression in linear response theory
\begin{equation}
\rho(\mathbf{r}, t) = \frac{1}{i\hbar} \int_{-\infty}^{t} dt' \left\langle \left[\rho(\mathbf{r}, t), H_{\rm J\cdot l}(t') \right] \right\rangle,
\end{equation}
which leads to the following expression in terms of the response function \(\boldsymbol{\phi}(\mathbf{r}, t)\)
\begin{equation}
\rho(\mathbf{r}, t) = \int d\mathbf{r}' \int_{-\infty}^{\infty} dt' \, \boldsymbol{\phi}(\mathbf{r} - \mathbf{r}', t - t') \cdot\mathbf{A}(\mathbf{r}', t').
\end{equation}

The charge-current response function \(\boldsymbol{\phi}(\mathbf{r} - \mathbf{r}', t - t')\) is defined as

\begin{equation}
\boldsymbol{\phi}(\mathbf{r} - \mathbf{r}', t - t') = \frac{\theta(t - t')}{-i\hbar} \left\langle \left[\rho(\mathbf{r}, t), \mathbf{j}(\mathbf{r}', t') \right] \right\rangle,
\end{equation}
where \(\theta(t - t')\) is the Heaviside step function. In the RPA, the function \(\boldsymbol{\phi}(\mathbf{r})\) is computed as

\begin{equation}
\boldsymbol{\phi}(\mathbf{r}) = \frac{e^2\hbar}{m} \sum_{\mathbf{k}, \mathbf{q}} e^{i (\mathbf{q} - \mathbf{k}) \cdot \mathbf{r}} \frac{\mathbf{k} + \mathbf{q}}{2} \frac{f_{\mathbf{k}} - f_{\mathbf{q}}}{\varepsilon_{\mathbf{q}} - \varepsilon_{\mathbf{k}}},
\end{equation}
where \(f_{\mathbf{k}}\) and \(\varepsilon_{\mathbf{k}}\) are the Fermi-Dirac distribution and the energy of the state \(\mathbf{k}\), respectively. Introducing the changes of variables \(\mathbf{k} = \mathbf{p} + \delta \mathbf{k}\) and \(\mathbf{q} = \mathbf{p} + \delta \mathbf{q}\), this expression becomes

\begin{align}
\boldsymbol{\phi}(\mathbf{r}) =& \frac{e^2\hbar}{m} \sum_{\delta \mathbf{k}, \delta \mathbf{q}} e^{i (\delta \mathbf{q} - \delta \mathbf{k}) \cdot \mathbf{r}} \left( \mathbf{p} + \frac{\delta \mathbf{k} + \delta \mathbf{q}}{2} \right) \frac{f_{\mathbf{p} + \delta \mathbf{k}} - f_{\mathbf{p} + \delta \mathbf{q}}}{\varepsilon_{\mathbf{p} + \delta \mathbf{q}} - \varepsilon_{\mathbf{p} + \delta \mathbf{k}}}\notag\\
=& \mathbf{p} \frac{e^2 \hbar}{m} \chi(r).
\end{align}
This reveals that \(\boldsymbol{\phi}(\mathbf{r})\) is proportional to the spin response function \(\chi(\mathbf{r})\) in Eq.~\ref{Eq.chiS}.

For the static charge density \(\rho(\mathbf{r})\), we then obtain the expression

\begin{align}
\rho(\mathbf{r}) =& (2 - g) e \hbar \mathcal{F} \int d\mathbf{r}' \, \mathbf{p} \chi(\mathbf{r} - \mathbf{r}') \nabla_{\mathbf{r}'} \times \mathbf{J} \delta^3(\mathbf{r}')
\end{align}
can be written in terms of the electric polarization \(\mathbf{P}, \rho=-\nabla\cdot \textbf{P}\) induced by the localized moment \(\mathbf{J}\)

\begin{equation}
\mathbf{P} = (2 - g) e \hbar \mathcal{F} \chi(\mathbf{r}) \mathbf{p} \times \mathbf{J}.
\end{equation}
This suggests that the spin density and the induced electric polarization are closely linked, with \(\mathbf{P}\) being proportional to the spin response function \(\chi\). This connection indicates that the charge density oscillates around the localized moment, similar to the spin density. The dependence on the factors \(\mathcal{F}\) and \((2 - g)\mathbf{J}\) emphasizes the critical role of skew scattering and orbital angular momentum in enabling the magnetoelectric coupling.

In the next section, we will delve further into the magnetoelectric coupling and discuss the exchange bias arising from the induced electric polarization.

\section{Results and Discussion} \label{SecResult}

\begin{figure}
\includegraphics[width=\columnwidth]{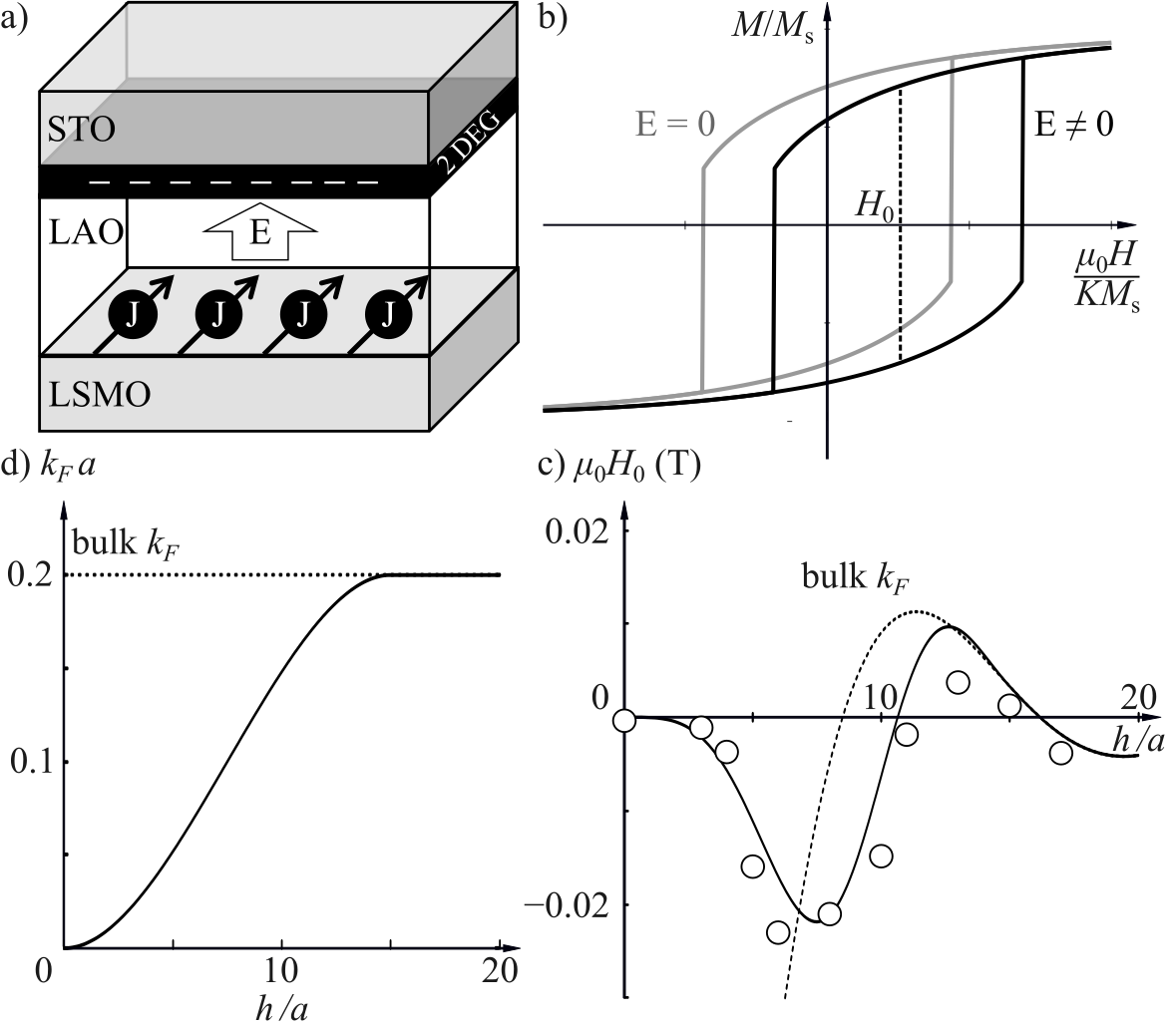} \caption{(a) Spin susceptibility that governs the indirect exchange interaction and exchange bias in the magnetic structure with quasi-one-dimensional symmetry. (b) A shift in the magnetic hysteresis loop, i.e., exchange bias, depends on the relative directions of magnetization $\textbf{M}$, electric field $\textbf{E}$, and the peak of the valence band $\textbf{p}$. 
{(c) exchange bias of LSMO$|$LAO$|$STO as a function of LAO width. Data points is taken from Ref.~\cite{Lu2016}. (d) Fermi wavevector $k_F$ is assumed to have smooth step from zero, when there is no spacer, to bulk value. The dotted line is for a constant $k_F$.}} \label{Fig.EB} 
\end{figure}

We now return to the system under investigation, which is the LSMO/LAO/STO magnetic heterostructure, as illustrated in Fig.~\ref{Fig.EB}a. In this system, the magnetic moments of LSMO can be modeled as a two-dimensional array located at positions $\textbf{r}_n$ on the $xy$-plane. This can be approximated as:

\begin{align} \textbf{J}\sum_n \delta^3(\textbf{r}-\textbf{r}_n) \approx N\textbf{J}\delta(z), \end{align} where $N$ represents the number of magnetic moments per unit area. In the following subsections, we will describe the phenomena mediated by spin interactions and the electric polarization of the LAO layer.

\subsection{Indirect Exchange Interaction Mediated by Itinerant Spin} \label{Sec:ExInteraction}

The effective magnetic field is localized at the $z=0$ interface and is given by:

\begin{align} 
\textbf{B} =& \frac{(g-1)\mathcal{I}}{\mu_B} \textbf{J} \sum_n \delta^3(\textbf{r}-\textbf{r}_n) \approx \frac{(g-1)\mathcal{I}N}{\mu_B} \textbf{J} \delta(z). 
\end{align}
As a result, the spin density depends only on the normal distance from the interface, $z$ (see App.~\ref{App:susc} for derivation), and is expressed as

\begin{align} 
\textbf{s} = (g-1)\mathcal{I}N\textbf{J} \chi_{q1D}(z), 
\end{align} 
where $\chi_{q1D}(z)$ is the spin susceptibility for the heterostructure with a quasi-one-dimensional system, given by
{
\begin{align} 
\chi_{q1D}(z) =& \frac{mk_F^2}{2\hbar^2\pi^2} f(k_Fz)\notag\\
f(x)=& \mathrm{Si}(2x) - \frac{\pi}{2}+\frac{2x \cos(2x) - \sin(2x)}{(2x)^2}
\end{align} 
}
where $\mathrm{Si}(x)$ is the sine integral function. This expression governs the spatial behavior of spin susceptibility in the heterostructure.

The $z$-dependent spin density enables indirect interactions across magnetic layers\cite{Yafet1987}. These interactions can be observed when the STO layer is replaced by another LSMO layer, forming a spin-valve structure \cite{Lu2016}. The corresponding Hamiltonian for the indirect exchange interaction is

\begin{align} 
H_\mathrm{indirect} =& -\int dz  \textbf{s}(z) \cdot (g-1)\mathcal{I}N \textbf{J}(z) \delta(z-h) \notag \\
=&-\mathcal{I}_\mathrm{RKKY}  \textbf{J}(0) \cdot \textbf{J}(h)\notag\\ 
\mathcal{I}_\mathrm{RKKY}=& \left( (g-1)\mathcal{I}N \right)^2 \chi_\mathrm{q1D}(h), 
\end{align} 
where $h$ is the thickness of the LAO layer. The interlayer exchange interaction, $\mathcal{I}_\mathrm{RKKY}$, exhibits a $2k_F z$ spatial oscillation characteristic of the RKKY interaction.

\subsection{Exchange Bias Mediated by Electric Polarization} \label{Sec:ExBias}

Similarly to the spin interaction, the effective vector potential due to the electric polarization is localized at $z=0$ and is given by

\begin{align} 
\textbf{A} \approx \frac{(2-g)m\mathcal{F}N}{e\hbar} \hat{z} \times \textbf{J} \frac{\partial \delta(z)}{\partial z}. 
\end{align}
The electric polarization, which depends only on the distance to the interface $z$, is expressed as
\begin{align} 
\textbf{P}(z) = (2-g) e \hbar \mathcal{F} N \textbf{p} \times \textbf{J} \chi_{q1D}(z). \label{Eq.P}
\end{align}

In addition to the localized magnetic field and vector potential from Mn, an out-of-plane electric field exists due to the two-dimensional electron gas (2DEG) at the LAO/STO interfaces. This electric field is modeled as:

\begin{align} 
\textbf{E}(z) = \hat{\textbf{z}} \frac{\sigma}{2\epsilon_0} e^{-\kappa |z-h|} 
, \label{Eq.E}
\end{align} 
where { $\sigma$ is charge density per unit area, $\kappa 
$} is the screening constant for the electric field \cite{Resta1977}. 

Substituting Eqs.~\ref{Eq.P} and \ref{Eq.E} into the electric energy yields the magnetoelectric coupling between the electric field and the magnetization, where the magnetization $\textbf{M}$ is given by

\begin{align} 
H_\mathrm{me} = -\int dV \ \textbf{P}(z) \cdot \textbf{E}(z) = V \textbf{M} \cdot \mu_0 \textbf{H}_0. 
\end{align} 
{Here parameter $\textbf{H}_0$ is given by

\begin{align} 
\mu_0\textbf{H}_0 
=& B_0\hat{\textbf{p}} \times \hat{\textbf{z}} \int_0^{h/a} dx e^{\kappa(ax-h)} f(k_Fax),\\
B_0=& \frac{(2-g) me Jk_F^2 \sigma \mathcal{F} NAa}{4\pi^2\hbar\epsilon_0M_s V} \approx 1 \textrm{ T} \notag
\end{align} 
The value of $B_0$ is estimated using $\sigma=e/(2a^2)$ \cite{Asmara2014}, $N=1/a^2$, $a=3.9$ \AA, $A=5$ mm $\times \ 3 \ \mu$m and $V=A\times10$ nm, $M_\mathrm{Mn}=2.2\mu_B$ \cite{Lu2016}, $M_s=2.81\times 10^4$A/m \cite{Jain2013,MatProj}, $m=92m_e$, $\mathcal{I}=2.14$ eV, $\mathcal{F}=\mathcal{I}/(4k_F)^2$, $g=1.8$, $k_F=0.2/a$, $\kappa=7.5/a$.
} 
The effect of this magnetoelectric energy on the hysteresis loop can be understood by incorporating $H_\mathrm{me}$ into the magnetic anisotropy energy, with the easy-axis-type anisotropy constant $K$

\begin{align} 
H_m =& V K \sin^2(\varphi-\alpha) - V \mu_0 \textbf{M} \cdot \textbf{H} + H_\mathrm{me} \notag \\ =& V K \sin^2(\varphi-\alpha) - V \mu_0 \textbf{M} \cdot (\textbf{H} - \textbf{H}_0), 
\end{align} 
where $\varphi$ is the angle of the magnetization $\textbf{M}$ relative to the magnetic field $\textbf{H}$, and $\alpha$ is the angle between the field $\textbf{H}$ and the easy axis. The hysteresis curve shown in Fig.\ref{Fig.EB}b is modeled using the Stoner-Wohlfarth model for $\varphi = 45^\circ$. This shift in the hysteresis loop center, caused by $\textbf{H}_0$, generates an exchange bias in the magnetic heterostructure. {$f(k_Fax)$ in the integration for $\textbf{H}_0$  generates a spatial oscillation of the exchange bias, as observed in Ref.[\onlinecite{Lu2016}]. The agreement of the experiment and our theory is illustrated in Fig \ref{Fig.EB}c. }

Finally, it is important to note that although centrosymmetric LAO implies that its valence band has two peaks at $R$ and $-R$, the magnetoelectric energy $H_\mathrm{me}$ will selectively enhance one peak and suppress the other. Thus, only the peak with empty states mediates interlayer exchange interactions. Consequently, the theory of a single valence band remains valid. The observed change in the exchange bias direction when the cooling field is reversed \cite{Lu2016} can be explained by shifts in the position of the valence band peak.

\section{Conclusion}
\label{SecSummary}

The results presented here provide a detailed analysis of the interaction between the localized Mn moments in LSMO and the itinerant charge and spin carriers in LAO using linear response theory. The main finding of this study is the identification of two primary mechanisms-exchange interaction and skew scattering-that govern the behavior of spin and charge densities in response to localized moments at the LSMO$|$LAO interface. These mechanisms lead to distinct spatial modulations of the spin and electric polarizations, which have important implications for the design of next-generation spintronic devices.

The exchange interaction between the localized moments and itinerant spins results in an oscillatory spin density around the localized moments, with a spatial frequency determined by the Fermi wave vector $k_F$. This oscillatory behavior reflects a complex interplay between the electron's spin and the localized moment, indicating that spin textures may be manipulated by tuning the strength of exchange interactions.

The skew scattering mechanism induces a charge density response, which is related to the orbital angular momentum of the localized moments. This leads to a spatial modulation of the charge density that mirrors the spin density oscillations. Additionally, the induced electric polarization, which is proportional to the spin response function $\chi$, underscores the magnetoelectric coupling between the electric and spin polarizations. The interplay between these two densities could be leveraged in the development of devices that integrate both magnetic and electric functionality.

The coupled nature of the charge, spin, and electric polarization in these heterostructures is crucial for the design of next-generation spintronic devices. The ability to control the spin and charge densities through localized magnetic moments and skew scattering interactions could enable the development of devices that combine both spin and charge transport, offering pathways for faster, more energy-efficient data processing. Additionally, the induced magnetoelectric coupling suggests potential applications in multiferroic devices, where the manipulation of magnetic properties through electric fields or vice versa could be exploited for non-volatile memory storage and logic operations.

In summary, the theoretical framework and results presented here highlight the complex interactions at the interface of LSMO and LAO, providing valuable insights for advancing the design of future spintronic and magnetoelectric devices. The control over charge-spin coupling and the emergence of induced electric polarization open exciting possibilities for creating multifunctional materials and devices that can harness both magnetic and electric effects for novel applications.

\begin{acknowledgments}
We thank Universitas Indonesia for funding this research through PUTI Grant No. NKB-366/UN2.RST/HKP.05.00/2024.
\end{acknowledgments}


%

\appendix
\section{Formalism and evaluation of susceptibilities}
\label{App:susc}

\subsection{Spin-spin Susceptibility}

The static spin susceptibility
\begin{align}
\chi_{ij}(\textbf{r})=&\sum_{\textbf{kq}}e^{i\textbf{q}\cdot\textbf{r}}
\frac{f_{\textbf{k}}-f_{\textbf{k}+\textbf{q}}}{\varepsilon_{\textbf{k}+\textbf{q}}-\varepsilon_\textbf{k}}\equiv\delta_{ij}\chi.
\end{align}
can be evaluated for two cases: (a) three dimensional system with a single impurity and (b) magnetic heterostructure with quasi-one dimensional symmetry.

\subsubsection*{Three Dimension}

\begin{align}
\chi&_\mathrm{3D}(\textbf{r})
=\iint \frac{d\textbf{k}d\textbf{q}'}{(2\pi)^6} e^{i\left(\textbf{q}'-\textbf{k}\right)\cdot\textbf{r}}
\frac{f_{\textbf{k}}-f_{\textbf{q}'}}{\varepsilon_{\textbf{q}'}-\varepsilon_\textbf{k}}\notag\\
=&\int \frac{d\delta\textbf{k} f_{\textbf{p}+\delta\textbf{k}}}{(2\pi)^3} \int \frac{d\delta\textbf{q}}{(2\pi)^3} 
\frac{e^{i\left(\delta\textbf{q}-\delta\textbf{k}\right)\cdot\textbf{r}}+e^{-i\left(\delta\textbf{q}-\delta\textbf{k}\right)\cdot\textbf{r}}}{\varepsilon_{\textbf{p}+\delta\textbf{q}}-\varepsilon_{\textbf{p}+\delta\textbf{k}}}\notag\\
=&\int_0^{k_F} \frac{4\pi(\delta k)^2d\delta k 2}{(2\pi)^3}\frac{\sin r\delta k}{r\delta k} \int_0^\infty  \frac{\frac{4\pi(\delta q)^2d\delta q}{(2\pi)^3} \frac{\sin r\delta q}{r\delta q}2}{\frac{\hbar^2}{2m}\left(-\left(\delta\textbf{q}\right)^2+\left(\delta\textbf{k}\right)^2\right)}\notag\\
=&-\frac{m}{\hbar^2\pi^3r^2}\int_0^{k_F} (\delta k)^2d\delta k \frac{\sin r\delta k}{\delta k} \cos r\delta k\notag\\
=&\frac{m}{8\hbar^2\pi^3}\frac{2k_Fr\cos(2k_Fr)-\sin (2k_Fr)}{r^4} 
\end{align}

\subsubsection*{Quasi One Dimension}
In quasi one dimension, $\textbf{B}(\textbf{r})=N(g-1)\mathcal{I}\mu_B^{-1}\textbf{J}\delta(z)$, $N$ is number of magnetic moments per unit area. In this case, integration for $\textbf{q}$ is only taken over $q_z$
\begin{align}
\chi_\mathrm{q1D}(z)
=&\int_{-\infty}^\infty \frac{dq}{2\pi} e^{iqz}
\int \frac{d\textbf{k}}{(2\pi)^3}\frac{f_{\textbf{k}}-f_{\textbf{k}+q\hat{\textbf{z}}}}{\varepsilon_{\textbf{k}+q\hat{\textbf{z}}}-\varepsilon_\textbf{k}}\notag\\
=&\int_{-\infty}^\infty \frac{dq}{2\pi} \cos qz
\int 4\pi r^2dr\frac{\sin qr}{qr}\chi_\mathrm{3D}(r)\notag\\
=&\int_z^\infty 2\pi r dr\chi_\mathrm{3D}(r)\notag\\
=&\frac{mk_F^2}{2\hbar^2\pi^2}\left(\mathrm{Si}(2k_Fz)-\frac{\pi}{2}\right.\notag\\
&\left.+\frac{2k_Fz\cos(2k_Fz)-\sin(2k_Fz)}{(2k_Fz)^2}\right)
\end{align}

\subsection{Charge-current Density}

On the other hand, the interaction $H_{\rm J\cdot l}$ can be written as an effective vector potential \begin{align}
\textbf{A}(\textbf{r})=&\frac{(2-g)m\mathcal{F}}{e\hbar}\nabla\times\textbf{J}\delta^3(\textbf{r}).
\end{align}
 acting on current 
\begin{align}
\textbf{j}(\textbf{r})=&\frac{e\hbar}{m}\sum_{\textbf{kq}\alpha} e^{i\textbf{q}\cdot\textbf{r}} \left(\textbf{k}+\frac{\textbf{q}}{2}\right) a_{\textbf{k}+\textbf{q}\alpha}^\dagger a_{\textbf{k}\alpha}
\end{align}
as can be seen below
\begin{align*}
H&_{\rm J\cdot l}=-\int d\textbf{r}\ \textbf{j}(\textbf{r})\cdot\textbf{A}(\textbf{r})\\
=&-\sum_{ijk}\epsilon_{ijk}\int d\textbf{r}\ \sum_{\textbf{kq}\alpha} e^{i\textbf{q}\cdot\textbf{r}} \left(\textbf{k}+\frac{\textbf{q}}{2}\right)_i a_{\textbf{k}+\textbf{q}\alpha}^\dagger a_{\textbf{k}\alpha}\\
&(2-g)\mathcal{F}\nabla_j\left(J_k\delta^3(\textbf{r})\right)\\
=&\sum_{ijk}\epsilon_{ijk}\int d\textbf{r}\ \sum_{\textbf{kq}\alpha} \nabla_j\left(e^{i\textbf{q}\cdot\textbf{r}}\right) \left(\textbf{k}+\frac{\textbf{q}}{2}\right)_i a_{\textbf{k}+\textbf{q}\alpha}^\dagger a_{\textbf{k}\alpha}\\
&(2-g)\mathcal{F}J_k\delta^3(\textbf{r})\\
=& -i\mathcal{F}\sum_{\textbf{kq}\alpha}(2-g)\textbf{J}\cdot\left(\textbf{q}\times\textbf{k}\right) a_{\textbf{k}\alpha}^\dagger a_{\textbf{q}\alpha}.
\end{align*}

The response of charge density 

\begin{align}
\rho\left(\textbf{r},t\right) =& \frac{1}{i\hbar}\int_{-\infty}^t dt' \left<\left[\rho\left(\textbf{r},t\right),H_\mathrm{\textbf{J}\cdot\textbf{l}}\left(t'\right)\right]\right>,\notag\\
=&  \int d\textbf{r}' \int_{-\infty}^\infty dt' \boldsymbol{\phi}\left(\textbf{r}-\textbf{r}',t-t'\right) \textbf{A}(\textbf{r}',t')
\end{align}
can be written in term of charge-current response function.
\begin{align}
\boldsymbol{\phi}\left(\textbf{r}-\textbf{r}',t-t'\right)=& \frac{\theta(t-t')}{-i\hbar}\left<\left[\rho\left(\textbf{r},t\right),\textbf{j}\left(\textbf{r}',t'\right)\right]\right>.
\end{align}
$\boldsymbol{\phi}$ can be found using similar procedure as spin-spin response  
\begin{align}
\boldsymbol{\phi}\left(\textbf{r}\right)
=&\frac{e^2\hbar}{m}\sum_{\textbf{kq}'}e^{i(\textbf{q}'-\textbf{k})\cdot\textbf{r}}\frac{\textbf{k}+\textbf{q}'}{2}
\frac{f_{\textbf{k}}-f_{\textbf{q}'}}{\varepsilon_{\textbf{q}'}-\varepsilon_\textbf{k}}
\end{align} 
setting $\textbf{k}=\textbf{p}+\delta\textbf{k}$ and $\textbf{q}'=\textbf{p}+\delta\textbf{q}$, one can see that 
\begin{align}
\boldsymbol{\phi}\left(\textbf{r}\right)=&
\frac{e^2\hbar}{m}\sum_{\delta\textbf{k}\delta\textbf{q}}e^{i(\delta\textbf{q}-\delta\textbf{k})\cdot\textbf{r}}\left(\textbf{p}+\frac{\delta\textbf{k}+\delta\textbf{q}}{2}\right)
\frac{f_{\textbf{p}+\delta\textbf{k}}-f_{\textbf{p}+\delta\textbf{q}}}{\varepsilon_{\textbf{p}+\delta\textbf{q}}-\varepsilon_{\textbf{p}+\delta\textbf{k}}}\notag\\
=&\textbf{p}\frac{e^2\hbar}{m}\chi(r)
\end{align} 


The static charge density for an impurity in three dimensional system 
\begin{align}
\rho\left(\textbf{r},t\right) =& (2-g)e\hbar\mathcal{F}\int d\textbf{r}' \textbf{p}\chi_\mathrm{3D}\left(\textbf{r}-\textbf{r}'\right) \nabla_{\textbf{r}'}\times\textbf{J}\delta^3(\textbf{r}')\notag\\
=& (2-g)e\hbar\mathcal{F}\int d\textbf{r}' \textbf{J}\delta^3(\textbf{r}')\cdot \left(\nabla_{\textbf{r}'}\times\textbf{p}\chi_\mathrm{3D}\left(\textbf{r}-\textbf{r}'\right) \right)\notag\\
\equiv& -\nabla\cdot \textbf{P}
\end{align}
can be written in term of electric polarization 
\begin{align}
\textbf{P}=(2-g)e\hbar\mathcal{F} \chi_\mathrm{3D}\left(\textbf{r}\right)\textbf{p}\times\textbf{J}
\end{align}


In quasi one dimension, 

\begin{align}
\textbf{A}(z)=&\frac{(2-g)mN\mathcal{F}}{e\hbar}\nabla\times\textbf{J}\delta(z).
\end{align}
The charge density and electric polarization are
\begin{align}
\rho\left(\textbf{r},t\right) 
=&-(2-g)e\hbar N\mathcal{F} \textbf{J}\cdot \left(\nabla\times\textbf{p}\chi_\mathrm{q1D}\left(z\right) \right),\notag\\
\textbf{P}=&(2-g)e\hbar N\mathcal{F} \chi_\mathrm{q1D}\left(z\right)\textbf{p}\times\textbf{J}.
\end{align}

\end{document}